\documentclass[a4paper]{CMNIEN}

\renewcommand\refname{REFERENCES}
\usepackage{graphicx}
  \usepackage{amsmath}
    \usepackage{amssymb}
    \usepackage{bm}
    \usepackage{amstext}
    \usepackage{amsthm}
    
\usepackage{float}
\makeatletter
\let\@tmp\@xfloat     
\usepackage{fixltx2e}
\let\@xfloat\@tmp                    
\makeatother

\title{THE ACTIAS SYSTEM: SUPERVISED MULTI-STRATEGY LEARNING
PARADIGM USING CATEGORICAL LOGIC}

\author{Carlos Leandro$^{\bf {1*}}$, Helder Pita$^{\bf 2}$, Lu\'is Monteiro$^{\bf 3}$}

\heading{THE ACTIAS SYSTEM}

\address{1: Departamento de Matem\'{a}tica\\
Instituto Superior de Engenharia de Lisboa, \\
Lisboa, Portugal.\\
e-mail: {miguel.melro.leandro@gmail.com}\\
\
\\
2: Departamento de Engenharia Electr\'onica de Telecomunica\c{c}\~oes e de Computadores,\\
Instituto Superior de Engenharia de Lisboa,\\
Lisboa, Portugal.\\
e-mail: {hp@isel.ipl.pt}
\
\\
3: Departamento de Inform\'atica,\\
Faculdade de Ci\^encias e Tecnologia, Universidade Nova de Lisboa,\\
Lisboa, Portugal.\\
e-mail: {lm@di.fct.unl.pt}
}


\abstract{One of the most difficult problems in the
development of intelligent systems is the construction of
the underlying knowledge base. As a consequence, the rate
of progress in the development of this type of system is
directly related to the speed with which knowledge bases
can be assembled, and on its quality. We attempt to solve
the knowledge acquisition problem, for a Business
Information System, developing a supervised multi-strategy learning paradigm. This paradigm is centred on a
collaborative data mining strategy, where groups of
experts collaborate using data-mining process on the
supervised acquisition of new knowledge extracted from
heterogeneous machine learning data models.
The Actias system is our approach to this paradigm. It
is the result of applying the graphic logic based language
of sketches to knowledge integration. The system is a data
mining collaborative workplace, where the Information
System knowledge base is an algebraic structure. It results
from the integration of background knowledge with new
insights extracted from data models, generated for specific
data modelling tasks, and represented as rules using the
sketches language}

\begin{document}
%\maketitle

\section{INTRODUCTION}
  The computational branch of theoretical computer
science is a well-established mathematical discipline
currently transforming itself into an applied category
theory. In contrast, data modelling is more closely
connected with practical information technologies. The
data modelling theory is not well formed
mathematically and is supported, outside the scope of
the relational data model, by ad hoc Database (DB)
theories. In the context of deterministic data modelling
this problem has been attacked with new and powerful
tools imported from category theory like sketches.
Sketches were proposed by Charles Ehresmann in the
late sixties, for the purpose of specifying algebraic
structures [21]. This graphic-based structure has a
mathematical foundation and offers a formal semantic
data-specification mechanism of quite different nature if
compared with the classical and informal mechanisms
like Chen's Entity-Relationship (ER) or Unified
Modelling Language (UML) but powerful to formalize
rigorously many modern data modelling problems
[16][4]. In the last decade there has been considerable
use of sketches to support semantic data modelling.
Important steps for its use were the development by
Piessens and Steegmans of an algorithm to verify the
equivalence of sketch models [16], Diskin design of
graphic query languages on sketches [4] and the
Michael Johnson and collaborators consultancy work
using sketches to Information System (IS) specification
and view integration [5] [9][10].
The core of an IS is one or a net of DBs. In the
modern view, DB presents an integral model of the real
world fragment (Universe of Discourse) upon which an
information system is built. A crucial step for the
information system design, business and data
understanding is to specify the universe in a consistent
and rigorous way. Which is the ideal field for applying
sketches, since they specify the universe in rigorous,
abstract and formalized terms using a graphical
language. It compresses the information structure in a
compact and comprehensible way, suitable for
communication among database experts and data
analysts.
The Universe of Discourse (UoD) specification using
sketches correspond to the specification of a set of
deterministic real-word constraints. This specification is
supposed to be valid during the IS life cycle. However,
in spite of its information changes, some relevant
probabilistic structures or patterns on its data can be
invariant, given useful and non-trivial knowledge about
the business rules and the data probabilistic structure.
The extraction of this type of information from large
repositories of information requires the use of automatic
techniques, which have being developed by Machine
Learning (ML) specialists. These techniques usually
generate probabilistic models, which generalise the data
structure, and traditionally are defined by different
semantic structures like neural networks, regression
trees, classification trees, rules sets, logical programs,
etc. The selection of one type of model depends from
the data structure and from the modelling task. This
selection is an important step of a process, called Data
Mining (DM) process, which has as goal the discovery
of knowledge. The knowledge discovery from data is
the result of an exploratory problem (e.g. [18])
involving the application of various algorithmic
procedures for manipulating data, building models from
data, and manipulating the models [6]. To define a DM
process involves many possible choices for each stage
of the process, which are domain dependent, if useful
information is found with a particular DM process on a
subset of the IS structure, probably it should have good
results in a new problem with similar data structure,
with minor change on the process (e.g. [1] and [2]). We
use the IS sketch specification as support for DM
process specification. This allows the definition of
libraries relating UoD structures with useful DM
processes to help solving new knowledge extraction
problems.
An airport, hospital or bank are discrete
heterogeneous systems consisting of many subsystems
of different kinds all integrated in a single whole via
complex mutual relationships and mutually dependent
functions subjected to specific business rules. Each
system can include its own specification language used
by the domain experts. Thus, the entire conceptual
model appears as a heterogeneous structure. The sketch
domain independence reduces the conceptual model
complexity, since it is sufficiently expressive to capture
the particularities of the whole system. Helping on the
development of collaborative data-mining strategies,
where teams of people with different profiles
collaborate on the IS knowledge extraction, flowing
different strategies and data views. Producing sets of
models for the same or different modelling tasks with
valuable new insights about the system information.
This knowledge, to be useful must be stored on a
knowledge repository, and deployed. To do this, and
because the semantic structure of data models can be
different from task to task or from strategy to strategy,
the model Integration requires a semantic language able
to present data models. The sketches seem to be helpful
in this context too.
The main goal of this work is study the advantages of
using sketches to represent information about the UoD,
IS structure and probabilistic data patterns and store it
on a knowledge database. This database can be seen as a
``big'' sketch describing the absolute specification of the
known business structure. This structure can be used to
induce new knowledge applying inference over it [4].
The new and helpful information can be used to enrich
the system sketch specification structure, with new
information, useful on the definition of new DM
extraction processes, and which can be propagated to all
the knowledge extraction workgroups.
The use of sketches allows the definition of what we
mean by supervised multi-strategy learning paradigm
The paper is organized as follows. In Section 2 the
Actias system architecture is presented as a set of
methodologies, each described in Sections 5, 6, 7, 8 and
9, and integrated through a business knowledge
repository. In Section 3 we begin with the definition of
basic concepts used on sketch data modelling and is
presented an example illustrating the category theoretic
information system specification techniques. Section 4
explains what we mean by collaborative data mining
and how the system is used for supervised multistrategy learning. Finally, in Section 10 we point out
future work on sketch data modeling on multi-strategy
learning.
\section{THE ACTIAS SYSTEM STRUCTURE OVERVIEW}
The Actias system supervised multi-strategy learning
goals and requisites are supported through a set of
methodologies. These methodologies take into account
the collaborative nature of the system. It classifies the
users in function of their jobs and profiles in a data
mining team, having in consideration the users
requirements for data analysis and modeling, DM
process specifications or data mining tasks definition.
Depending on the user profile he sees the system from
different perspectives corresponding each one to an
operational view of the system. This decomposition
extends the CRISP-DM specification, having as metagoals the creation and\\or administration of a business
knowledge repository, the heart of the system.
The Business Knowledge repository is defined by
four types of business information: (1) Structural
Models built using sketches and that correspond to the
business model with UoD, IS warehouse specification
and probabilistic patterns; (2) Data Stream Processes
defined by DM extraction processes; (3) Data Models
defined by heterogeneous model structures each
associated with a type of learner; (4) Normalized Data
Models corresponding to the presentation of information
extracted from the models and presented as fuzzy sketch
predicates.
Functionally the Actias system can be decomposed in
a (see Figure 1):
\begin{figure}[h]
\begin{center}
\includegraphics[width=350pt]{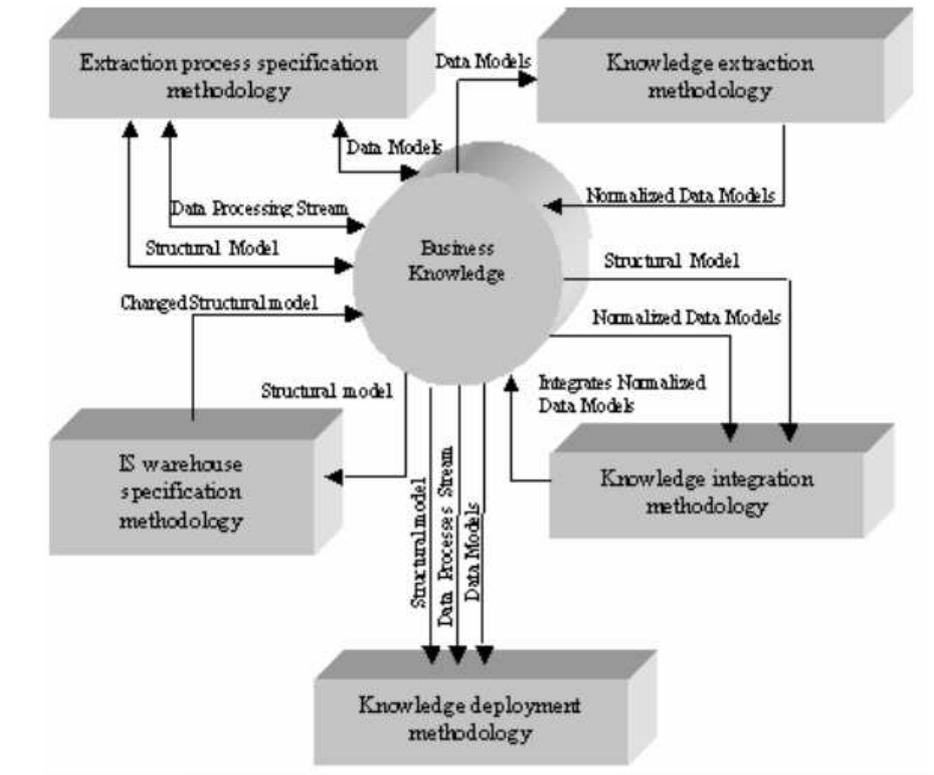}
\end{center}
\caption{The Actias system structure}\label{fig1}
\end{figure}
\begin{itemize}
\item \textbf{IS Warehouse specification methodology}; where
the language used to describe the business domain
and its structure is specified (or edited) using a
domain hierarchical decomposition strategy. This
decomposition is based on libraries with sketch
specification background knowledge and primitive
structures. This methodology can have as input the
meta-information associated with the IS warehouse
and its output is a sketch specifying the
deterministic information structure:
\item \textbf{Extraction process specification methodology},
corresponding to a set of functions and processes
supported by a graph language used to specify or
edit DM processes to modelling a data stream,
define by sketch queries, based on the WEKA
toolkit [20] for data processing algorithms, and
organized on a CRISP-DM process taxonomy.
\item \textbf{Knowledge extraction methodology}, where the
relevant new insights are extracted from the data
models and its representation is normalized using
sketches.
\item\textbf{ Knowledge integration methodology}, where the
new knowledge is integrated in the business sketch
on the knowledge repository, preserving the
consistency.
\item \textbf{Knowledge
deployment
methodology},
corresponding to models or business knowledge
deployment to data users or decision support
teams
\end{itemize}
The relations between these methodologies and the
business knowledge repository are presented in figure
bellow and are described at the subsequent sections.

% Figura 1 the Actias System Structure
\section{DATA MODELLING USING SKETCHES}
In this section we provide the background for the
sketch data model. We assume some familiarity with
ER models and with some elementary category theory.
Any introduction to category theory (for example [21])
contains the basic category theory definitions needed
below including commutative diagrams, limits, and
coproducts.
Following [9] a sketch is a directed graph whose
nodes specify entities and attributes, similar to an ER
diagram, and a set of categorical constraints defined on
this graph. These constraints take three forms:
\begin{itemize}
\item \textbf{Commutative diagrams}, defined by pairs of paths
in the graph with common origin and destination.
\item \textbf{Limit constraints}, specifying a certain node in the
graph as the ``Limit'' of a specified diagram in the
graph.
\item \textbf{Colimit constraints}, specifying a certain node in the
graph as the ``Colimit'' of a specified diagram in the
graph.
\end{itemize}
For practical reasons usually restrictions are
imposed on the limit and colimit constrains. The type of
this categorical constraints allowed on a sketch
specification determines the sketch semantic power.
Which can be used to define a hierarchy of sketches
ranging from sketches with very little expressive power
but well-behaved model categories easy to use for data
specification, to sketches with stronger expressive
power but less well-behaved model categories and more
hard to use and understand (e.g. [17]). For our needs, we
limit data sketch specification to finite limit constraints
and coproducts. Coproduct constraints specify that a
certain node in the graph is to act as the ``coproduct'' of
specified nodes in the graph. This restriction simplifies
the users specification tool. The semantic loss imposed
by this restriction doesn't affect its usability. This type
of sketch has enough power to define structures
specifiable using Horn-clauses, which correspond to the
type of information structures present in Information
Systems [19].

An Information system, sometimes called a database
state or instance, is an assignment of, for every node the
sketch, a finite set of instances or values of that entity or
attribute, and for every arrow in the sketch graph a
relation between the corresponding entity instances,
such that:
\begin{itemize}
\item The commuting diagrams do indeed commute as
diagrams of the corresponding relations.
\item The sets assigned to limit nodes are limits of the
corresponding diagrams of relations.
\item The sets assigned to coproduct nodes are indeed
coproducts, i.e. disjoint unions, of the assigned
sets.
\end{itemize}
In other words, a database state is a diagram of sets
and relations, which have the same shape as the sketch
graph, and whose sets and relations satisfy the
constraints.

A business data model is essentially a mathematical
structure, is intended to correspond in more or less
detail to the business activity and information,
representing the business understanding. Like
mathematical semantics, data semantics arises in two
forms, which might be described as absolute semantics
and relative semantics. In absolute semantics terms, the
meaning of an object is fully determined by its
interactions, by the operation it can perform and the
properties which they satisfy. Relative semantics, on
other hand, permits a hierarchical structuring of
semantics by allowing a new structure to be defined
based on another. Business data sketches are essentially
absolute semantic structures but its definition can follow
a relative methodology. This methodology is used to
reduce the specification task complexity. It allows the
hierarchy domain decomposition where more complex
structures are defined based on pre-existent ones, which
allows the definition of libraries with useful structures.

To let the use of sketches as relative semantic
structures the system expresses the sketch syntax using
boxes and arrows, context diagrams and diagram marks.
Context diagrams provide a format for representing
models graphically and allowing hierarchal domain
problem decomposition. A sketch is defined by a set of
diagrams, which can cross-reference each other. Each
graphic diagram contains boxes, arrows, box/arrow
interconnections and associated relationships. Where
boxes represent each major object of the domain, such
as entities, data sources, or data sets. These objects can
be decomposed into more detailed diagrams, until the
domain is described at a level necessary to support the
goal of a particular data-modelling task. The top-level
diagram provides the most general or abstract data
structure represented by the model. This decomposition
represents a refinement process ending on a set of
primitive objects corresponding to primitive business
entities. Parallel to the object decomposition, the system
allows a similar process for arrows. An arrow on the
child diagram can be seen as refining a meta-arrow
defined on the parent diagram.

The specification of the business IS using a sketch
corresponds to a rigorous way to present the business
understanding and its data structure understanding. This
knowledge is important for the definition of data mining
tasks and specify the knowledge extraction processes.
However, the business knowledge is dynamic changing
with time, which imposes its enrichment with new
knowledge. Given by domain experts or extracted from
symbolic models generated by machine learning
algorithms and usually presentable by first-order many-sorted logic formulae. The connection between sketch-
based logic and classical logic has been studied
exhaustively (e.g. [13]), giving us methods to translate
logic theories to sketch theories and vice-versa. We are
making progresses studying its application to fuzzy
logic. This will allows us to do the normalization of the
expert knowledge and data knowledge with the business
model, allowing the business sketch data enrichment
with new insights.

\section{COLLABORATIVE DATA-MINING}
The uses of sketches for data modelling has been
criticized for being too complicated, especially
concerning its formalism and the way of thinking about
the specification process which is different from the one
used on the standard methods. At the beginning, its use
may seem to be overkilled. However, once the sketch
methodologies have been understood, this structure can
quickly be applied to lots of different data extraction
problems with some domain independence. Helping on
the development of collaborative data-mining
strategies, as a way to interchange business knowledge
among people with different profiles and expertises.

Knowledge discovery from data can be seen as
teamwork, where different experts collaborate on the
business and data exploration process specification.
Which involves the application of various procedures
for storing and integrating data on a repository,
specifying the data structure, manipulating data,
building models from data, manipulating the models,
extracting relevant information from models, integrating
and deploying knowledge. Which impose the user
specialisation on the use of some algorithms and
techniques. To incorporate a collaborative work strategy
the system distinguishes between three user types:
Information System Administrator, Data Analyst, and
Knowledge Engineer. The Information System
Administrator specifies the business information system
warehouse using sketches, changes it, and implements a
data security policy defining the user work groups and
groups data views. The Data Analysts follow a CRISP-
DM methodology [18] to specify and deploy knowledge
extraction processes using IDEF0 diagrams, based on a
data view, and implements a problem security policy
defining projects, tasks and task data views. The
Knowledge Engineer tests the extraction processes
deployed by the Data Analyst and extracts relevant
information from the generated models, deploy it and
tries to integrate it on the IS sketch with the existing
domain knowledge without significant loss of its
consistence level. The Knowledge Engineer is also
responsible for defining the level of consistence for the
rules used on domain enrichment processes and for the
rule confidence update during the IS life.
\begin{figure}[h]
\begin{center}
\includegraphics[width=400pt]{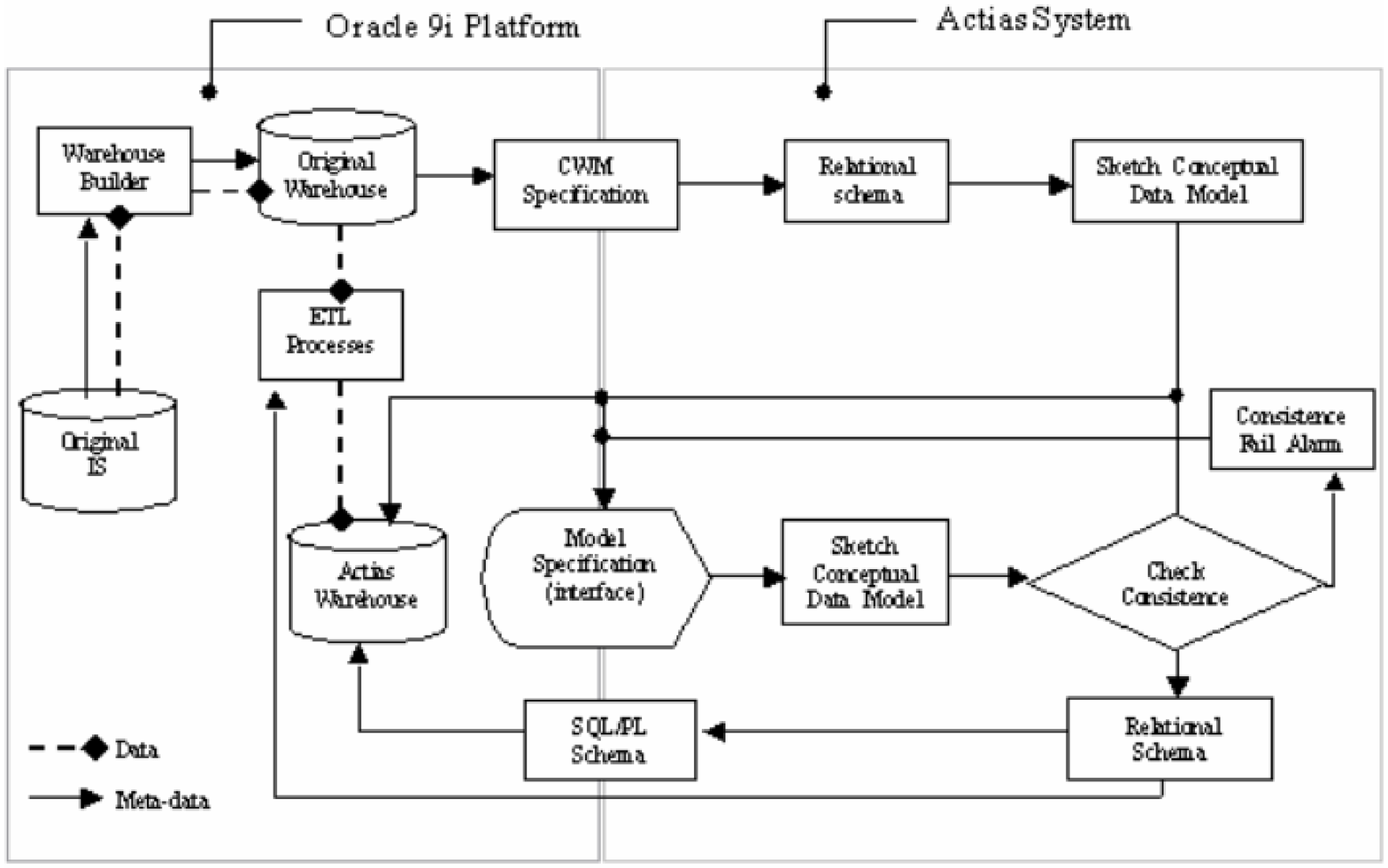}
\end{center}
\caption{Relation between the IS model and its original Data Warehouse.}\label{fig2}
\end{figure}

\section{INFORMATION MODELLING}
The Actias system information modelling
methodology has the goal of defining a business sketch
model based of the small part of the world associated
with the business Information System. This small part
of the world corresponds to the business universe of
discourse (UoD). The models of the data specification
must be seen as possible states of the UoD, and are the
information storable on its structures or equivalently
stored in the business IS. This process is incremental
following the Business Knowledge needs and evolution.
Therefore the problem of knowing whether the
specification and its changes are compatible with the
existent IS information structure become visible. This
requires the development of a strategy to validate the
model against the IS structure. For this we were
inspirited by the G. Karács and P. van Bammel
approach, where the IS structure is first mapped to an
intermediate specification. Based on the CWM
(Common Warehouse Meta-data) description a business
IS warehouse, we produce the IS representation using a
relational schema, see Figure 2, which can be seen as a
sketch [8], a first approximation to the business
conceptual model, helping the user on the specification
work. This relational schema will be used to verify if
the update of business model is able to store all the
information present on the IS. After performing the
information compatibility test, some times is
convenient, for operational reasons, to generate or
change the internal Actias system Data Warehouse,
improving its structure based on the new knowledge.
Helping on the definition of sketch views and data
streams for the extraction process specification
methodology.

The sketch structure definition or enrichment is done
using a CASE tool, implementing a refinement
hierarchic methodology allowing the gradual
introduction of grater levels of detail through the sketch
diagram structure.
Having as main goal the
minimization of potential specification complexity
based on a Modelling Knowledge library, containing
helpful structures. Producing as result a process to
generate or change the warehouse, reflecting directly the
structures specified by the enriched sketch.

The information sketch modelling methodology can
be decomposed, see Figure 3, into five generic phases:
Base relational model acquisition, sketch model
enrichment (using the CASE tool), implementation
sketch generation, sketch model validation and
warehouse structural revision using the implementation
sketch. The produced warehouse adaptation tries to
make easier its use on the extraction processes. Since, it
requires the definition of a data stream, which will be
seen as a query result, defined on the user sketch model
view selecting in it a part of the general sketch having
associated a data dictionary [11].

\begin{figure}[h]
\begin{center}
\includegraphics[width=400pt]{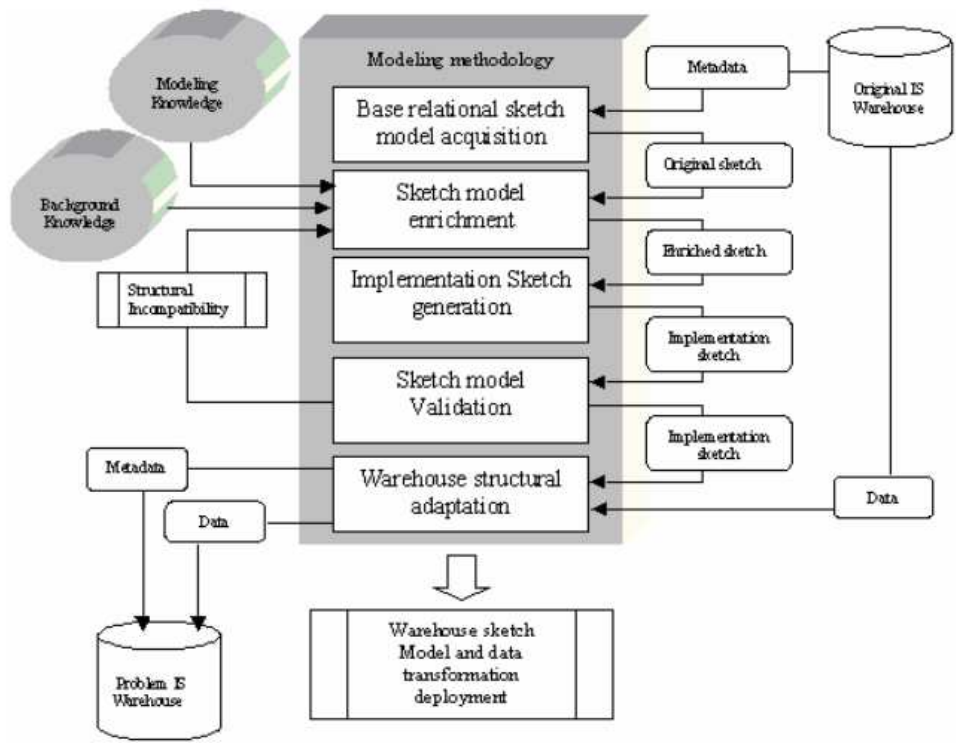}
\end{center}
\caption{Modeling methodology.}\label{fig3}
\end{figure}

\section{EXTRACTION PROCESS SPECIFICATION}
A knowledge extraction process involves multiple
stages. A simple, but typical, process might include
processing data, applying a data-mining algorithm, and
post-processing the mining results. There are many
possible choices for each stage, and only some
combinations are valid. The primary goal of the
extraction process methodology is to provide support to
this, based on data-mining and data processing
ontology.

As planning methodology we selected the SADT
(Structures Analysis and Design Technique) more
precisely the IDEF0 (Integration DEFinition language
0) [7]. This choice is motivated by the fact that it is a
graphic language, based on atomic building blocks,
provided with methods allowing: (1) a structured
decomposition of complex planning problems; (2) the
definition and management of transferred data between
(sub)problems, the definition of control variables and
the allocation of resources needed for its execution; (3)
record the decision made and the results; (4) make the
model evaluation about its completion, consistency and
correctness. One of the most important features of
IDEF0 as a modelling concept is that it gradually
introduces greater levels of detail through the diagram
structure comprising the model. These levels of detail
end with atomic functions defined based on a library of
primitive functionalities. Defined based on the WEKA
Data Mining library [20] for Java and organized
following a specific data-mining ontology.

The ontology contains for each operator [1]: a
specification of the condition under which the operation
can be applied, involving a precondition on the state of
the extraction process as well as its compatibility with
preceding operation; a specification of the operation?s
effects on the extraction process state, on the data, and
on the sketch which specify the domain structure;
logical group, which can be used to narrow the set of
operations to be considered at each stage of the
extraction process; predefined schemata for generic
problems indexed by a sketch structure associated with
its data stream; a help function to obtain online
information about each of the operations; a set of rules
to be used by an agent to check the extraction process
structure consistency.

Figure 4 shows a structural view of the ontology,
which groups the extraction operations into six major
groups: business understanding; data understanding;
pre-processing; data analysis and control; modelling;
and post-processing.

\begin{figure}[h]
\begin{center}
\includegraphics[width=450pt]{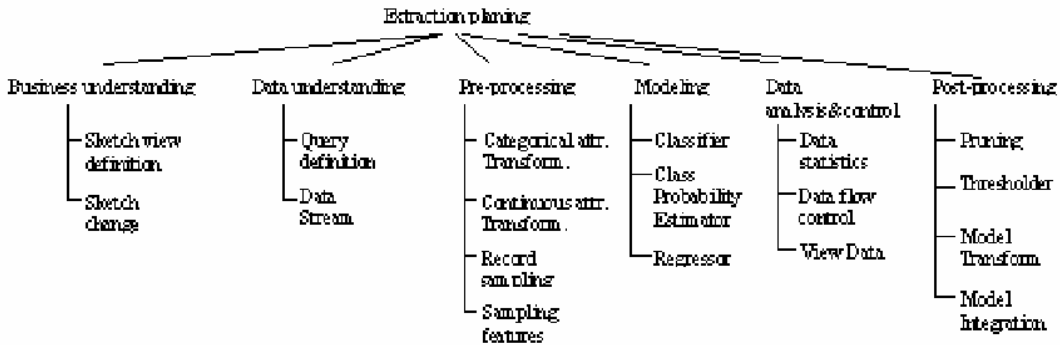}
\end{center}
\caption{Extraction process ontology.}\label{fig4}
\end{figure}

The Actias system uses an ontology-based approach
on a CASE tool to specify extraction processes using
IDEF0 diagrams, following the general framework
present on figure 5 [7]. Where the Background
Knowledge implements the task ontology, with access
supported by the Interaction mechanism, and used to
define or change existent extraction processes. It is
represented by the logic architecture reflecting the
problem refinement. Which, after functional resources
have been associated with its primitive tasks, generates
the Realisation architecture. If this architecture is valid,
i.e. satisfies the agent consistency check, it can be seen
as a solution to the data extraction task, and can be
executed, producing as result a model or a set of
models. On the process of producing this models the
extraction process changes the domain sketch
specification. These changes are consequence of the
possible use of auxiliary structures on the specification
process definition and used as on the resulting model
preconditions.

\begin{figure}[h]
\begin{center}
\includegraphics[width=450pt]{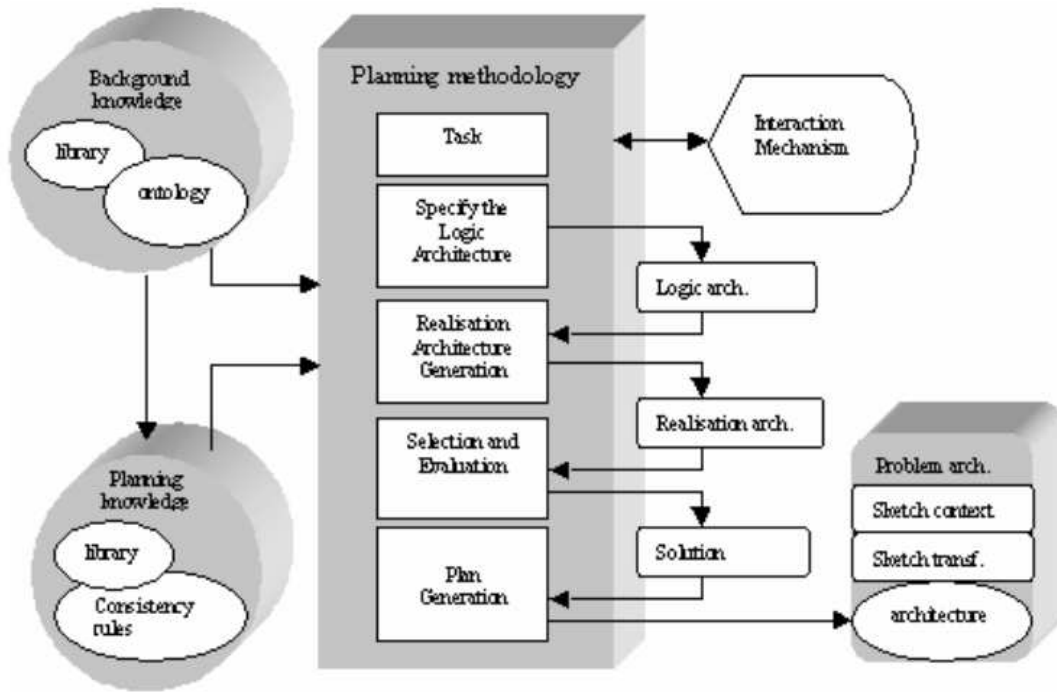}
\end{center}
\caption{Planning methodology.}\label{fig5}
\end{figure}

\section{KNOWLEDGE EXTRACTION}
After the execution of an extraction process the
generated model or set of models can provide new
insights on the business, which should be deployed to
the other users through the enrichment of the domain
sketch model. This requires the development of an
interface able to display rules extracted from models
and compute its relative importance against the
information stored on the Actias data warehouse. This
requires the models information normalization; since the
system uses different types of machine learning models.
For this we use probabilistic Horn-clauses. These types
of rules are easy to extract from symbolic models, and
when the model is connectionist, it is translated to a
symbolic one using a mining process.

The probabilistic Horn-clauses are similar to the way
human experts express their expertise and users are
comfortable with this way of expressing newly
extracted knowledge. This is important during the
expert validation of a knowledge base and for the
knowledge deployment.
Having an important advantage, they are easy to translate to the graphic
based logic of sketches.

\section{KNOWLEDGE INTEGRATION}
The knowledge integration methods allow the
combination of any collection of learned models, i.e.,
they may be generated by a collection of heterogeneous
or homogeneous data models. This integration can be
done on the extraction process as a post-processing
integration function, or through the knowledge
integration methodology if it is made on model
extracted Horn-clauses. The integration on extracted
rules is performed using its sketch language
representation, and only on rules selected by the
Knowledge Engineer. The use of sketch simplifies this
task, since the sketch model of a set of rules associated
with a data model can be seen as the model of a Horn-
theory. Then to integrate the sets of rules extracted
from, for instance, two models correspond to integrate
two Horn-theories, which have been extensively studied
for sketches, and can be performed algorithmically for
finite sketches [17]. And to produce the knowledge base
update with the new insights the integration process is
extended to all the business sketch specification.

This integration method has a reduce sensitivity to
the correlation in learning models because the learning
models are condensed into a set of uncorrelated basic
set of rules. Only those sets accounting for significant
niches are retained for integration on the business
sketch. The combining set of rules discarded the
redundancy in the set extracted from learning models by
discarding the rules, which do not account for a
significant amount of information established by the
Knowledge Engineer.

When new theories are integrated on the business
specification, its structure is enriched. This induces the
change of the user structure views, allowing the
definition of new data modelling tasks having as data
stream the result of queries defined on parts of this new
structure.

This phase closes the knowledge base enrichment
cycle. The relations among the types of information
involved in it are represented in Figure 6.

\begin{figure}[h]
\begin{center}
\includegraphics[width=450pt]{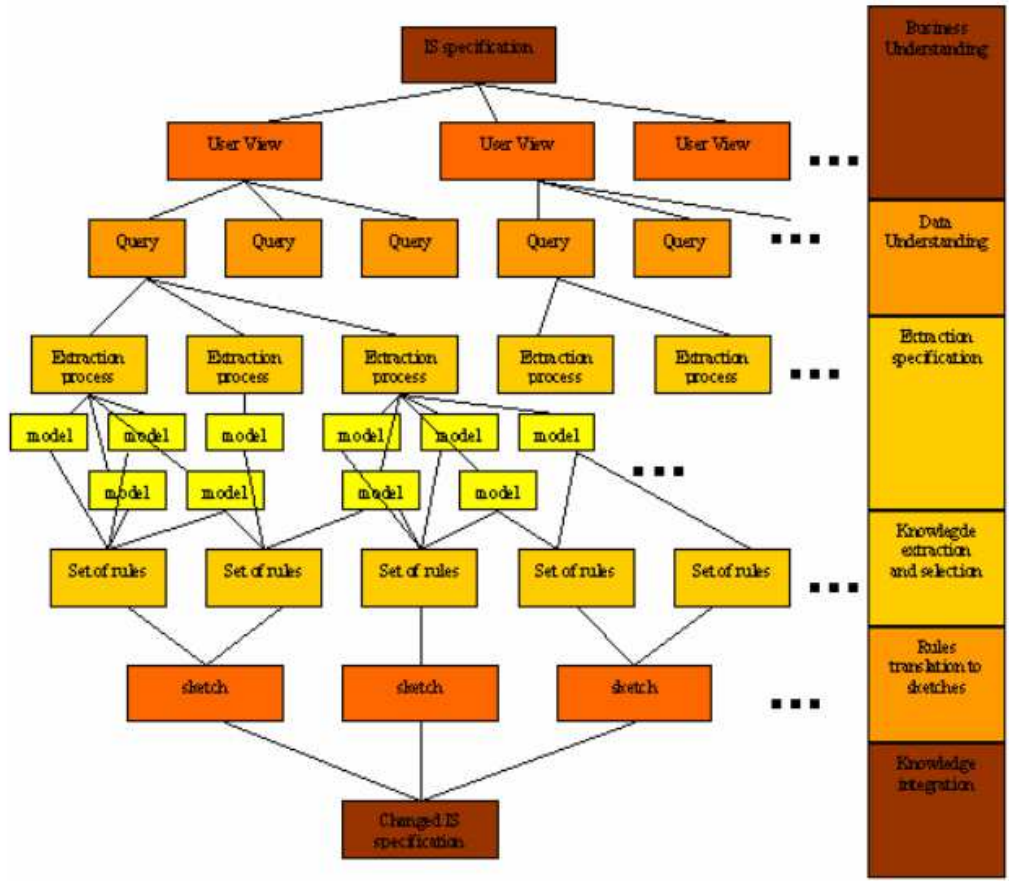}
\end{center}
\caption{Information structures and its relations on the Actias system.}\label{fig6}
\end{figure}

\section{KNOWLEDGE DEPLOYMENT}
The system uses a web-based interface for
knowledge deployment. Through it user accesses to
models, executes extraction processes and views the
business knowledge. When the interface is used to
access a model, depending on the type of model, a
prediction or a distribution is generated. The
consistency of a theory extracted from a model changes
during the IS lifetime, which requires the update of the
rules confidence used on the business specification. This
type of update is done through a knowledge deployment
interface, executing the required extraction process.

The database of rules used on the sketch business
definition is accessible and represents the business
knowledge base. This information can be used to define
intelligent systems using the business IS. This strategy
is used to simplify the sketch direct use. However the
direct sketch edition through the web interface was
considered, but it is only available to sketch data
modelling experts since it is difficult to read for non-
experts.

\section{CONCLUSIONS AND FUTURE WORK}
The Actias system is in the specification phase. Its
success depends on the integration level reached among
its modules. Its usability is highly conditioned by the
way the interfaces simplify the sketch specification
process and the query definition, and how this process is
integrated with the IDEF0 data processing specification
CASE tool. At present we are developing prototypes for
these interfaces.

This is the first of a series of articles exploring the
advantages of using the sketch formal language on the
specification of an Information System universe of
discourse. We are extending the results on deterministic
structure modelling using finite sketches to a
probabilistic framework. This will allow us to apply
sketch logic inference techniques on the business model.
Generating new probabilistic knowledge expressed
directly by sketches, which will extend the system
induction capabilities.

\section{ACKNOWLEDGMENTS}
This work is being supported in part by the research
project no 37/2003 from Instituto Polit\'ecnico de Lisboa.

\end{document}